\documentclass[twocolumn,aps,prl]{revtex4-2}
\usepackage[latin9]{inputenc}
\usepackage{amsmath}
\usepackage{amssymb}
\usepackage{graphicx}
\usepackage{ulem}
\usepackage{lineno}
\usepackage[colorlinks,linkcolor=blue,anchorcolor=blue,citecolor=blue,urlcolor=blue]{hyperref}
\makeatletter

\usepackage{color}
\usepackage{bm}
\usepackage{float}

\makeatother

\begin{document}
\title{
	Highly Polarized Energetic Electrons via Intense Laser-Irradiated Tailored Targets
}

\author{Xiaofei Shen}
\email{shen@mpi-hd.mpg.de}

\author{Zheng Gong}

\author{Karen Z. Hatsagortsyan}
\email{k.hatsagortsyan@mpi-hd.mpg.de}

\author{Christoph H. Keitel}

\affiliation{Max-Planck-Institut f\"ur Kernphysik, Saupfercheckweg 1, 69117 Heidelberg, Germany}

\date{\today}

\begin{abstract}

A  method for the generation of ultrarelativistic electron beams with high spin polarization is put forward, where a tightly-focused linearly-polarized ultraintense laser pulse interacts with a nonprepolarized transverse-size-tailored solid target. The radiative spin polarization and angular separation is facilitated by the 
standing wave formed via the incident and reflected laser pulses at the overdense plasma surface. Strong electron heating caused by transverse instability enhances photon emission in the density spikes injected into the standing wave near the surface. 
Two groups of electrons with opposite transverse polarization emerge, anti-aligned to the magnetic field, which are angularly separated in the standing wave due to the phase-matched oscillation of the magnetic field  and the vector potential. The polarized electrons propelled into the plasma slab, are focused at the exit by the self-generated quasistatic fields.
Our particle-in-cell simulations demonstrate the feasibility of 
highly polarized electrons with 
a single 10~PW laser beam, 
e.g. with polarization of 60$\%$ and charge of 8~pC selected at energy of 200~MeV within 15~mrad angle and 10$\%$ energy spread.

\end{abstract}

\maketitle

Spin-polarized electron beams are an essential tool for   for physical research with applications spanning from particle physics \cite{Pick2008,Accardi2016,Abbott2016} to material science \cite{Feder1986,Wolf2001,Zutic2004}. 
Conventionally, spin-polarized relativistic electron beams are mainly produced via extraction from a photocathode \cite{Pierce1976} and further acceleration 
\cite{Hartmann1990}. The alternative is radiative spin polarization (RSP) in storage rings 
\cite{Sokolov1964,Sokolov1968,Baier1967,Baier1972}. However, both methods require
unique and expensive large-scale accelerators, and the RSP in storage rings is rather slow. 
Recently, methods consisting of pre-polarizing a gas jet via photodissociation with circularly-polarized (CP) lasers  and then  accelerating them in a laser wakefield, were proposed \cite{Wen2019,Wu2019,Buscher,Fan2022}, but multiple laser beams are required and the electrons are depolarized in the injection process. Such ideas are more refined in Refs.~\cite{Nie2021,Nie2022} where polarized electrons are created during ionization injection of electrons
with a CP laser into a particle-driven  
wakefield accelerator, but how to extend this to laser wakefields remains unclear.

With the rapid development of laser technology, PW and 10 PW  laser facilities have been constructed worldwide such as  \cite{Yoon2021,ELI,SULF}, with a perspective for  100 PW  
\cite{Danson2019,Shao2020,XCELS,Gonoskov2022}. 
Such intense laser pulses have important applications for nonlinear QED \cite{Piazza2012,Gonoskov2022,Fedotov2023} including $\gamma$-ray \cite{Ji2014,Gonoskov2017,Sampath2021} and pair production \cite{Chang2015,Ridgers2012,Bell2008,Piazza2016,Vranic2017}. A question of growing interest is how to construct a laser-driven ``mini-storage-ring" to efficiently polarize electrons via RSP in fs timescale by using ultrastrong laser fields. An additional highly polarized electron
beam can significantly enhance research perspectives of a PW laser facility.
 Unfortunately, the basic concept of nonlinear Compton scattering in a strong monochromatic laser wave provides low  polarization because of the oscillating character of the laser field:
the polarization built up in a half-cycle is lost in the subsequent one \cite{Yu2004,Seipt2018,Kotkin2003,Karlovets2011}. To break the interaction symmetry  and produce electron beams with net high spin polarization, several schemes have been proposed, such as colliding  electron beams with elliptically-polarized \cite{Li2019} or two-color \cite{Seipt2019,Chen2019} pulses, and electrons trapped in the magnetic node of two colliding CP pulses \cite{Sorbo2017,Han2022,Qian2023}. However,  experimental realization of the aforementioned schemes is still challenging, because it requires precise manipulation of laser polarization \cite{Danson2019}, or precise spatial and temporal control over multiple laser and electron beams \cite{Poder2018,Cole2018}. An especially limiting factor is the pointing stability of the beams.

The RSP in laser-plasma interaction (LPI) has been also analyzed recently \cite{Song2022,Gong2021,Gong2023,Xue2023}. In \cite{Song2022}, polarized positron production in QED cascades is considered with a laser intensity $I_L\simeq3\times10^{24}\,{\rm W/cm^2}$. There, the  asymmetry for the  spin-flip in the propagating laser wave stems from the wave damping due to pair production, which allows to induce net polarization of about $20\%$. The role of  quasistatic magnetic fields in LPI for RSP has been investigated in Refs.~\cite{Gong2021,Gong2023,Xue2023} with an emphasis on applications for plasma diagnostics \cite{Gong2021,Gong2023}. 
The question of whether highly polarized dense relativistic electron beams could be obtained via intense LPI
still remains open.

\begin{figure}
	\includegraphics[width=8.2cm]{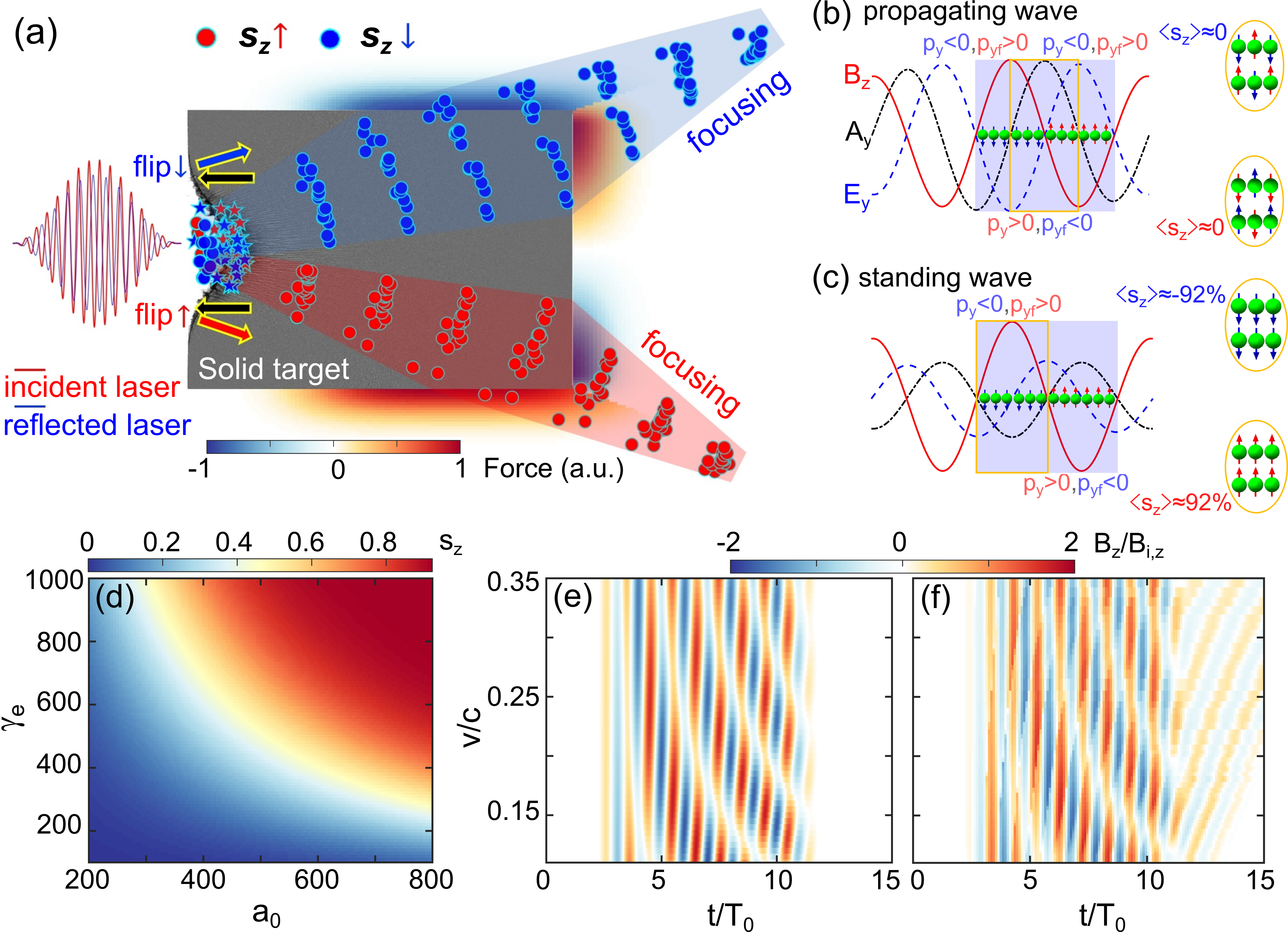}
	\caption{(a) Interaction setup: an intense
	laser pulse impinges on a transverse-size-tailored  solid target (gray). Counterpropagating electrons undergo spin-flip upon being pushed back into the target (arrows). Determined by instant local fields, the electrons have different divergences and are flipped to oppositely transverse polarization (pentagrams mark their initial positions). Upon exiting the target, a  focusing force is self-generated  confining the electrons. Illustration of the polarization and separation of electrons in a propagating (b) and standing (c) wave field, where $p_y$ is the transverse momentum when the flip occurs and $p_{yf}$ is the final momentum. (d) Theoretical estimation of $s_z$ in a half-cycle using Eq. (\ref{tau}). Temporal evolution of $B_z$ at $x=0$ for different hole-boring (HB) velocities: (e) theoretical prediction via Eq. (\ref{standing_wave}), (f) via PIC simulations with  parameters introduced in the text.
				}\label{fig:fig1}
\end{figure}

In this Letter, we investigate RSP of electrons via the interaction of an ultraintense tightly-focused linearly-polarized (LP) laser pulse with nonprepolarized overdense plasma, see Fig.~\ref{fig:fig1}(a), aiming at  highly polarized electrons  with a single 10~PW-class laser beam. 
Particle-in-cell (PIC) simulations have been carried out
with the EPOCH code \cite{Arber2015}, in which the radiative spin flips and spin precession  have been implemented \cite{Gong2021}.
In the setup, the interaction symmetry is deliberately  broken due to the reflection of the laser pulse at the overdense plasma boundary and the consequent formation of a standing wave.  The strong transverse instability  at the front surface \cite{Pegoraro2007,Shen2017} gives rise to density spikes that penetrate  into the stronger field  region at the surface within a quarter-wavelength, where electron heating and  photon emissions  induce strong net RSP of electrons. 
The electrons are propelled  into the plasma by light pressure without evident depolarization.   
They are split into two oppositely transversely-polarized parts because of the polarization  and  transverse momentum correlation  in the standing wave field [Fig. \ref{fig:fig1}(c)]. This  contrasts to the case of a propagating wave [Fig. \ref{fig:fig1}(b)], without transverse separation with respect to polarization. By
transverse-size-tailoring (TST) of the target,
the focusing of the polarized electrons at escaping the target  is achieved by the self-generated quasistatic field. Despite of a broad momentum distribution of outgoing electrons, still the selection can yield a dense highly polarized relativistic electron beam, e.g., with  the 10$\%$ energy and 15~mrad angular windows,  an electron beam  with $60\%$ polarization, and  8 pC charge  is obtained with a 10~PW laser.

We first consider a simplified model of our setup, where an ultraintense LP laser pulse, with vector potential $A_{i,y}=a_0{\rm sin}(x-t)$, impinges on an overdense target. Here,  $a_0$ is the  field amplitude, and dimensionless units are used: $t\rightarrow\omega_Lt$, $x\rightarrow x\omega_L/c$, $v\rightarrow v/c$, $A_y\rightarrow eA_y/m_ec^2$,
where $\omega_L$ is the laser frequency, while $m_e$ and $e$ are the electron mass and charge, respectively. The light pressure is known to induce HB effect,  with the characteristic
velocity $v_{\textrm{HB}} = \sqrt{\Pi}/(1+\sqrt{\Pi})$ \cite{Robinson2009},
where $\Pi = I_L/m_in_ic^3$, $m_i$ and $n_i$ are the ion mass and density, respectively. The reflected laser pulse in the HB frame is \cite{Macchi2009,Kostyukov2016}: $A_{r,y}'=a_0{\rm sin}(x'+t')$ (the primed variables belong to the moving frame),
forming a standing wave:
\begin{eqnarray}
&&A_y' = 2a_0\sin\left(x'\right)\cos(t'),\nonumber\\
&&B_z' = 2a_0\cos\left(x'\right)\cos(t'), E_y' = 2a_0\sin\left(x'\right)\sin(t').\;\;\;
	\label{standing_wave}
\end{eqnarray}
This would facilitate RSP: (i)
The amplitude of $B_z'$ is doubled, and it oscillates in phase with $A_y'$ [Fig. \ref{fig:fig1}(c)], different from the propagating wave [\ref{fig:fig1}(b)];  (ii) $E_y'$ peaking at $x'=-\pi/2$ would help confining electrons to a small region \cite{Supp}; 
(iii) The pulse is ``folded", leading to highly asymmetric interaction.

For RSP, $B_z$ is responsible, whose time evolution at $x=0$ for different $v_{\textrm{HB}}$ is shown in Fig.~\ref{fig:fig1}(e). The ``stripes" where the field is vanishing, originate from the time-varying $v_{\textrm{HB}}$
and have been verified by our  PIC simulations [Fig.~\ref{fig:fig1}(f)], where $v_{\rm HB}$ is manipulated via varying electron density $n_e$.  The slower $v_{\textrm{HB}}$ (higher $n_e$ for the same $I$), the fewer the ``stripes". Hence, employing high-density targets is beneficial for pursuing higher RSP.

Let us estimate RSP in our setup. Generally, at synchrotron motion, RSP is gradually built up according to  $s_z(t) = s_0\left(1 - e^{-t/\tau_0}\right)$ \cite{Baier1972},
with the characteristic polarization time $\tau_0=(8/5\sqrt{3})(m_e^2c^2/e^2\hbar)(R^3/\gamma_e^5)$, and the RSP upper limit  $s_0=8/5\sqrt{3}\approx0.9238$. Here,
$\hbar$ is the reduced Planck constant, $R$ the radius of the synchrotron motion, and $\gamma_e$ the electron Lorentz-factor. Applying the synchrotron RSP in our case in the frame moving  with $v_{\rm HB}$,
where the electron undergoes circular motion  with $R'=p'_\perp/(eB')$, we arrive at the RSP time:
\begin{eqnarray}
\tau_0\approx\frac{8}{5\sqrt{3}}\frac{m_e^5c^5}{e^5\hbar}\frac{\gamma^{3}{\rm sin}^3\theta_e}{\gamma_e^2B_z^3},
	\label{tau}
\end{eqnarray}
where $\gamma=1/\sqrt{1-v_{\textrm{HB}}^2}$ and $\theta_e={\rm atan}(p_\perp/p_x)$ is the electron divergence angle.  The RSP  in a half-cycle, estimated by Eq.~(\ref{tau}), which corresponds to the maximum  achievable polarization in LPI, is shown in  Fig.~\ref{fig:fig1}(d). The larger $\gamma_e$ ($a_0$), the higher $s_z$ expected. With $a_0=600$ and  $\gamma_e=1000$, $s_z$ reaches $91\%$, approaching the RSP limit of $92.38\%$ \cite{Baier1972}. Here we assume that electrons lose $50\%$ energies to photons and $\theta_e=60^\circ$  obtained from simulations with immobile ions.

\begin{figure}
	\includegraphics[width=8.2cm]{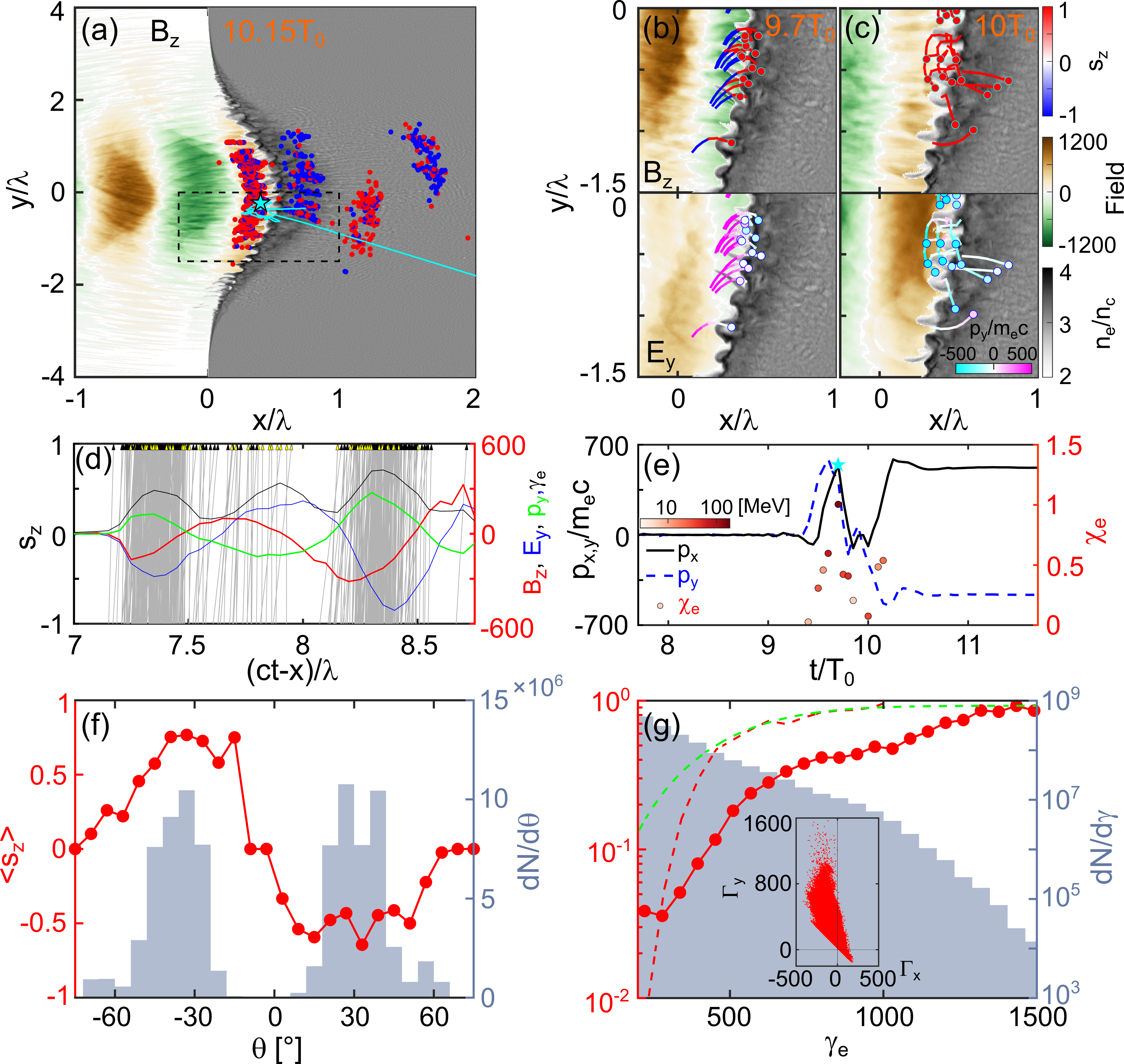}
	\caption{
		\label{fig:fig2}
		Generation of highly spin-polarized electrons: (a) The laser pulse (contour)  bores a hole on the target front (gray) when its peak arrives.
		The red (blue) dots present spin-up (down) electrons and the cyan line depicts a typical electron trajectory with details shown in (d) and (e).
		(b), (c) Snapshots to illustrate the correlation between electron emission direction and $s_z$, where the curves represent the electron trajectory within 0.3$T_0$ color-coded with their instant $s_z$ (upper panel) and $p_y$ (lower panel).
		(d) Evolution of $s_z$ (gray), averaged $p_y$ (green), $B_z$ (red) and $E_y$ (blue). The arrow color indicates the direction of final $p_{yf}$ (black for $p_{yf}<0$, yellow $p_{yf}>0$). Here only electrons with spin flipped up are shown.
		(e) Evolution of the electron momenta and $\chi_e$ color-coded with the photon energy. The asterisks mark where (a) and when (e) its spin is flipped. (f) Angular distribution of $\left<s_z\right>$ (red dots) and electron number (blue) at the source for $\gamma_e>1200$. (g) The corresponding $\left<s_z\right>$ and energy spectrum vs  $\gamma_e$ at around $-32^\circ$. The red dashed line shows  the case with immobile ions (see Supplemental Materials (SM) \cite{Supp}), while the green is the estimation using Eq.~(\ref{tau}). The inset illustrates the energy-gain plane ($\Gamma_x,\Gamma_y$) where $\Gamma_{x,y}=-\int ev_{x,y}E_{x,y}dt$.
	}
\end{figure}

The estimation above is further proven by two-dimensional PIC simulations with realistic parameters. We use a laser pulse with peak intensity of $5\times10^{23}\,{\rm W/cm^2}$ ($a_0=600$ for a laser wavelength of $\lambda_L=1\mu$m). The pulse is $y$-polarized with a transverse Gaussian profile and focal spot size $d_L=2\mu$m,
duration of $15$ fs with $\sin^2$ temporal profile (feasible in upcoming 10 PW-class lasers~\cite{Gonoskov2022,Danson2019}).
The initial phase is $\pi$ to optimize the spin polarization. A uniform rectangular target is considered with thickness of $9.4\mu$m and transverse size $10\mu$m to focus electrons at about $30^\circ$, and
$n_e=1200n_c$ to maintain the target opaque without
excessive computational efforts,
where $n_c=\pi m_ec^2/e^2\lambda_L^2$ is the critical density. The  restricted ion mobility is favorable, and considering the strong laser field, we have chosen the ion species to be Au$^{69+}_{197}$ according to the Perelomov-Popov-Terent'ev ionization model \cite{Perelomov1966,ADK}. The simulation box ($x,y$) is $17.4\mu$m$\times20\mu$m with resolution of $0.002\lambda_L\times0.008\lambda_L$. The macroparticles in each cell for electrons and ions are 1000 and 40, respectively. The pair production is neglected because their number is only 1$\%$ of high-energy electrons, and Bremsstrahlung plays a minor role for high-energy electrons traversing a thin foil \cite{Supp}. Demonstrations of the  robustness  of our results (regarding effects of prepulse, oblique incidence, pointing stability, temporal profile, initial carrier envelope phase and 3D effects)  as well as numerical convergence are presented in  SM \cite{Supp}.

When a laser pulse irradiates an overdense plasma,
transverse instabilities inevitably develop at the interface \cite{Pegoraro2007,Shen2017}, leading to the formation of density spikes, as shown in Fig.~\ref{fig:fig2}(a). These density spikes penetrate into stronger field region, where electrons can be pulled out and accelerated to higher energies via $\bf j\times B$ heating, which is confirmed by the dominance of the transverse energy gain, see inset in Fig. \ref{fig:fig2}(g) \cite{Gibbon2005,Supp}. The maximum electron energy exceeds 800~MeV [\ref{fig:fig2}(g)], much higher than the typical oscillating energy, even though radiation reaction (RR) takes away substantial energy. The electron quantum
strong-field parameter $\chi_{e}=(e\hbar/m_e^3c^4)|F_{\mu\nu}p^\nu|$ reaches about 1 [\ref{fig:fig2}(e)], with
the field tensor $F_{\mu\nu}$, and the electron four-momentum $p^\nu$. The high-energy electrons in the strong field region have larger probabilities for RSP.
Following the trajectory illustrated in Fig.~\ref{fig:fig2}(a) (cyan) and SM movie \cite{Supp}, most flips occur at the density spikes where the local field and $\chi_e$ are stronger [\ref{fig:fig2}(e)].

The picture of the electron emission direction correlated with the spin polarization [Fig.~\ref{fig:fig2}(a-c)] has a simple explanation. It originates from the phase-matched oscillation in time of  the magnetic field and the vector potential in a standing wave. The HB effect is minor for the polarization resolved angle separation (see the plane wave case in SM \cite{Supp}). In RSP,  $s_z$ tends to flip to the opposite direction of $B_z$ [upper panel in Fig.~\ref{fig:fig2}(b)], while $p_y$
has the same sign as $A_y$ (opposite to $B_z$), see Eq. (\ref{standing_wave}) and Fig.~\ref{fig:fig2}(d).  Accordingly, the electron polarization and  $p_y$ at the photon emission are correlated, and the latter, along with the photon recoil, determines  the final $p_{yf}$ after the fast injection into the plasma [Fig.~\ref{fig:fig2}(c)] \cite{Supp}. 
Consequently, in the plasma slab, the electrons with different polarization move in different transverse direction [Fig.~\ref{fig:fig2}(a)]. The angle between them   is sufficiently large [Fig.~\ref{fig:fig2}(f)] to separate them in an experiment. We underline the difference from the propagating wave case [Fig.~\ref{fig:fig1}(b)], where for the given $B_z$, both signs of $p_y$ are possible, hindering the separation over spin. Note that in our setup, the generation of hot electrons near the overdense target front and their injection into the plasma is an ultrafast process ($<1T_0$), during  which each electron experiences only a single spin-flip and the spin depolarization is suppressed. This is because the electron energy is strongly reduced after the photon emission and the subsequent fields are much weaker [Fig.~\ref{fig:fig2}(e)]. This facilitates approaching the maximal theoretical RSP limit.

The RSP is created in a small region ($<\lambda_L/4$) ahead of the HB front, and
we evaluate the mean spin polarization $\left<s_z\right>$ created in this region (polarization at the source) in Fig.~\ref{fig:fig2}(g). Here,
$\theta_e=-32^\circ$, where the electron polarization is highest and at $t=14T_0$, when the RSP ceases.
One can see that after the photon emission and injected into the plasma, the high polarization $\left<s_z\right>$ is still correlated with the high $\gamma_e$, similar to the theoretical prediction. The reason is that, when bounced back into the plasma, the electrons regain their initial energy [Fig. \ref{fig:fig2}(e)].
With $\gamma_e\sim1000$ (1500), $\left<s_z\right>$ reaches about 50$\%$ ($88\%$), approaching the upper limit of RSP.
This is significantly higher than previous LPI schemes \cite{Gong2021,Song2022}. Nevertheless, at the same $\gamma_e$, the polarization in the simulation is slightly lower than the theoretical limit [green in Fig.~\ref{fig:fig2}(g)]. This is
because of the curved HB front. Electrons from the wings originate with moderate energies  and stay in the laser field for longer time after the spin-flip, leading to slight depolarization.

  \begin{figure}[b]
	\includegraphics[width=8.2cm]{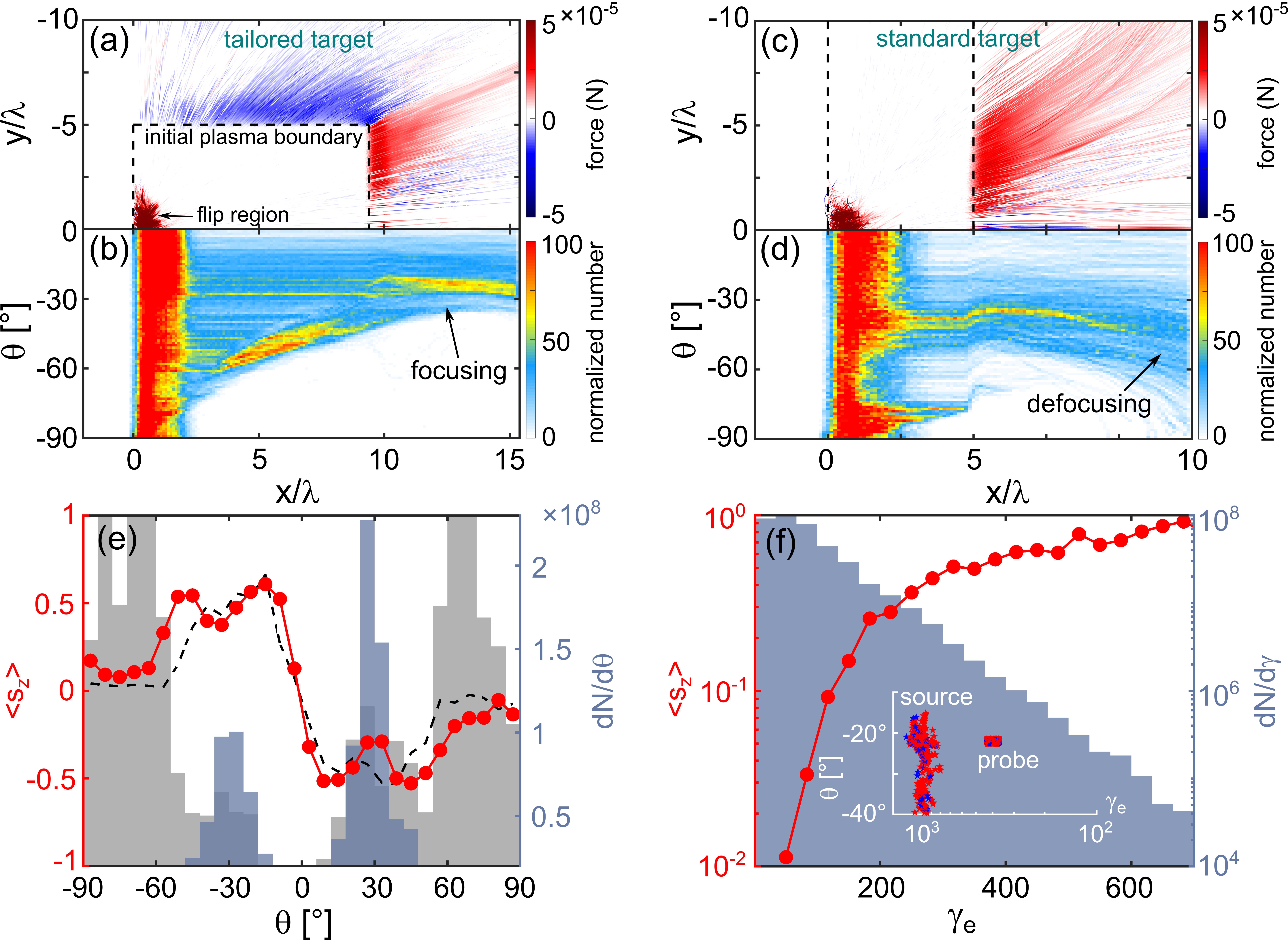}
	\caption{
		\label{fig:fig3}
		Focus and collection of polarized electrons. (a) Trajectories of high-energy electrons with final $\theta_e<0^\circ$ color-coded with the instant force felt by them, where blue (red) represents an inward (outward) scattering force. (b) $\theta_e$ vs $x$, where the colormap represents the relative electron number. (c,d) correspondingly show the results from the comparison case with a standard target. In (a,c), black dashed lines mark initial plasma boundaries. (e) Angular distributions of $\left<s_z\right>$ (red for TST, black for comparison case) and electron number (blue for TST, gray for comparison case)
		for electrons collected at the simulation boundaries
		with $\gamma_e>300$. (f) The corresponding $\left<s_z\right>$ and energy spectrum vs $\gamma_e$. The inset displays the focusing of selected electrons with $s_z$ color-coded.
	}
\end{figure}

For applications the polarization of the outgoing electron beam is important. Therefore, extracting the polarized electrons is crucial. For the standard transverse-size-unlimited target, outgoing electrons  mainly experience longitudinal deceleration due to the sheath field established at the target rear \cite{Macchi2013,Shen2021}, which induces strong defocusing and greatly reduces the electron number at the angle of peak polarization, see Figs.~\ref{fig:fig3}(c-e).

The advantage of using the TST target is the focusing effect  for the outgoing electrons due to the electrostatic fields induced at the target surface, as schematically illustrated in Fig.~\ref{fig:fig1}(a). This provides a new degree of freedom for manipulating the electron beam. Such targets are in use in current LPI experiments \cite{Buffechoux2010}.
Figure~\ref{fig:fig3}(a) illustrates the force acting on the electrons traversing the target boundaries,  where blue (red) indicates an inward (outward) scattering force. This is determined by comparing the instant $\theta_e$ with the surface force direction $\theta_f$. It is clearly shown that at the corner, electrons are focused from both sides to around $30^\circ$
[Fig.~\ref{fig:fig3}(b)].

\begin{figure}
	\includegraphics[width=8.2cm]{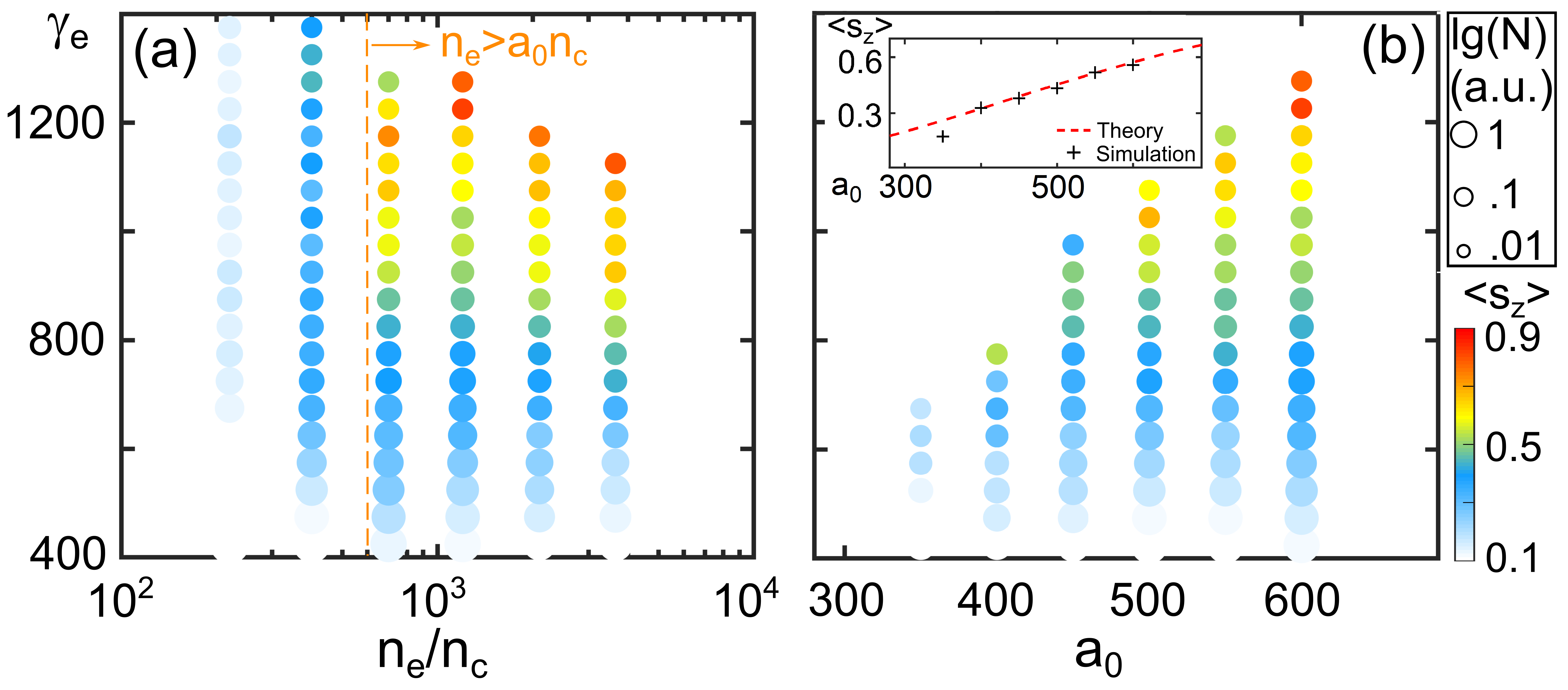}
	\caption{
		\label{fig:fig4}
		Scaling. (a), (b) Dependence of $\left<s_z\right>$ on $n_e$ (same $a_0=600$) and $a_0$ (same $n_e=1200n_c$). The marker size represents the charge. (c) Scaling of $\left<s_z\right>$ with respect to $a_0$, with $\gamma_e>1.5a_0$: simulation (black crosses) and theory (red).
	}
\end{figure}
The final polarization distributions of outgoing electrons collected at the simulation boundaries depending on $\theta_e$ and $\gamma_e$  are shown in Figs.~\ref{fig:fig3}(e) and  \ref{fig:fig3}(f). The electron number peaks at about $\pm30^\circ$ with narrower opening angles than at the source. By selecting $\gamma_e>200$, $400$ and $700$, $\left<s_z\right>$ is about $40\%$, $60\%$ and $90\%$, respectively.
Thus, the TST target helps to focus electrons, but  has  no side effects on the polarization. The minor impact of the laser beam jitter on the focusing properties of TST target is discussed in SM \cite{Supp}. 
In plasma-based electron acceleration, the number of low-energy electrons always largely exceeds that of high-energy ones regardless of acceleration mechanisms, but they are of little interest and cannot be polarized limited by Eq. (\ref{tau}). Thus, beam transport lines  have been widely employed in laser wakefield acceleration to filter them out  and improve the beam parameters \cite{Andre2018,Tilborg2015,Maier2020}. By passing through the beam transport line  developed in \cite{Andre2018}, 
it is feasible to select electrons at 200~MeV within 15~mrad and $10\%$ energy spread [inset of Fig.~\ref{fig:fig3}(f)] without much loss. The beam charge still reaches approximately 8~pC and the polarization is about $60\%$. Note that without selection the total charge exceeds 110~pC at  polarization $60\%$, and 2~pC at $90\%$. 
Compared to the standard case, the electron charge is about twice higher [Fig.~\ref{fig:fig3}(e)] and can be further enhanced by optimizing the target parameters.

The role of the plasma ($n_e$) and laser  ($a_0$)  parameters on RSP is analyzed in Figs.~\ref{fig:fig4}(a), (b).
For $a_0=600$ and relativistically-overdense plasma ($n_e>a_0n_c$), the physical process is similar and the maximum polarization can always reach above $80\%$ at the high-energy part. However, for $n_e<a_0n_c$, both the time and space scales of LPI are extended, and due to much weaker reflection, the interaction symmetry cannot be broken. Hence, the final polarization  decreases, vanishing at $n_e\ll a_0n_c$.

In Fig.~\ref{fig:fig4}(b), $\left<s_z\right>$ increases with $a_0$ and $\gamma_e$, following the prediction of Eq.~(\ref{tau}). We can provide analytical scaling for the average $\left<s_z\right>$ at $\gamma_e>1.5a_0$ and $a_0\in[350,600]$ discussed here. From simulations we found that the factor of $\gamma{\rm sin}\theta_e/B_z$ remains almost constant  because an increase in $a_0$ is accompanied by an increase of both $\gamma$ ($v_{\rm HB}$) and $\theta_e$ (caused by a deeper HB front). Consequently, the RSP time $\tau_0\propto1/\gamma_e^2$ and $P\approx P_0[1-e^{c_0(\gamma_e/100)^2}]$ with $c_0=-0.0121$ characterizing the polarization efficiency in a half-cycle. A good agreement of this  analytical estimation with the simulation results is illustrated in the inset of Fig.~\ref{fig:fig4}(b). It shows that with a weaker laser field of $a_0=400$ and higher energy selection $\gamma_e>2a_0$, the polarization can still reach about $50\%$.

Concluding, we have shown the feasibility of producing highly polarized dense relativistic electron beams by using an upcoming 10 PW-class laser interacting with TST targets. Applying additionally  a practical selection technique of the beam transport lines, high-energy low spread electron beams with polarization $>60\%$ and considerable charge are achievable, that are auspicious for extensive applications.


\end{document}